# Carbon-Doped Sulfur Hydrides as Room-Temperature Superconductors at 270 GPa

S. X. Hu,[1,2,*] R. Paul,[1,2] V. V. Karasiev,[1] and R. P. Dias[2]

[1]Laboratory for Laser Energetics, University of Rochester,

250 East River Road, Rochester, NY 14623-1299, USA

[2]Department of Mechanical Engineering, University of Rochester,

250 East River Road, Rochester, NY 14623-1299, USA

*Corresponding author. Email: shu@lle.rochester.edu

ABSTRACT

To understand the most-recent experiment on *room-temperature* superconductivity in carbonaceous sulfur hydride (CSH) systems under high pressure, we have performed extensive stoichiometry and structure searches of ternary CSH compounds using generic evolutionary algorithms. Judged from the formation enthalpy of different CSH compounds, our studies conclude that certain levels of carbon doping (~5% to 6%) in sulfur hydride ($H_3S$) in its *R3m* and *Im$\bar{3}$m* phases gives rise to the most-stable structure (second to the $H_3S$ itself) among the various CSH systems found. The replacement of a small amount of sulfur atoms by carbon in compounds like $C_1S_{15}H_{48}$ and $C_1S_{17}H_{54}$ results in a stronger electron−phonon coupling and a higher averaged phonon frequency that increases with pressure, thereby leading to room-temperature superconductivity at ~270 GPa. The calculated superconducting transition temperature $T_c$ of $C_1S_{15}H_{48}$ and $C_1S_{17}H_{54}$ as a function of pressure shows reasonably good agreement with experimental measurements. Before transition to superconducting states at ~80 GPa, the CSH system is predicted to have a stoichiometry of $C_2S_2H_{10}$ with a stable structure of P1 symmetry, which is supported by the direct comparison of its Raman spectrum with experiment.

Superconductivity is a physics phenomenon occurring in certain materials that shows no resistivity for electrical current at temperatures below a critical transition temperature $T_c$. For conventional superconductors, the electron–phonon coupling mediates the formation of a Cooper pair of electrons with opposite spins, offering a physics picture to understand superconductivity by the Bardeen–Cooper–Schrieffer (BCS) theory [1]. Extensive studies on conventional superconductors had continued for decades after the establishment of the BCS theory, with a wide range of superconducting elements and alloys discovered. However, the highest transition temperature any conventional superconductors reached so far was $T_c$ = 39 K in $MgB_2$ [2]. The search for high-temperature superconductors has shifted since the discovery of superconductivity in cuprate perovskite materials [3] in the 1980s, with a current record of $T_c$ = 138 K at ambient pressure [4] ($T_c$ = 157 K at 23.5 GPa pressure [5] and $T_c$ = 164 K at 31 GPa pressure [6]) is still being held by mercury–barium–calcium–copper–oxide ($HgBa_2Ca_2Cu_3O_x$). In recent years, high-pressure–induced superconductivity in hydrogen-rich materials (so-called "hydrides") became a hot topic, following the pioneering suggestion of heavy atom doping in hydrogen to lower the required pressure for observing superconductivity [7].

For the past five years, exciting high-$T_c$ superconducting materials have been discovered from binary hydride or super-hydride systems of $X_mH_n$ in both experiments [8−12] and density-functional-theory (DFT) calculations [13−17]. In particular, the first experimental synthesis of $H_3S$ by compressing $H_2S$ gave an unprecedentedly high superconducting transition temperature of $T_c$ = 203 K at a pressure of $P \approx 150$ GPa [8]. The subsequent discoveries of heavy-metal super-hydrides have further pushed the superconducting temperatures to $T_c$ = 250 K at 170 GPa [10] and $T_c$ = 260 K at $P \approx$ 180 to 200 GPa [11] both for $LaH_{10}$, as well as the recent experimental observation [12] of $T_c$ = 263 K at $P$ = 200 GPa in $YH_9$. Most recently, a room-temperature superconductor has been experimentally realized in a carbonaceous sulfur hydride (CSH) system [18], which demonstrated a transition temperature of $T_c$ = 287.7±1.2 K at 267±10 GPa. This experimental finding has excited the condensed-matter and high-pressure physics communities to understand this "puzzling" ternary CSH system, as recent studies [19,20] prior to the experiment [18] did not predict room-temperature $T_c$ behavior.

In this Letter, we report on *first-principles* calculations and physics understanding of the stoichiometry and structure of the newly discovered room-temperature superconducting CSH system. Our results indicate that carbon-doped $H_3S$ in the *R3m* and *Im$\bar{3}$m* phases are energetically more stable than any other carbon-rich CSH compounds. A certain carbon-doping level (~5% to 6%) in $H_3S$, such as $C_1S_{15}H_{48}$ and $C_1S_{17}H_{54}$, gives rise to room-temperature superconductivity at ~270 GPa. The calculated superconducting transition temperature $T_c$ of $C_1S_{15}H_{48}$ and $C_1S_{17}H_{54}$ versus pressure has a positive slope in agreement with experimental measurements, which results from an enhanced electron–phonon coupling and higher average phonon frequency at higher pressures. At lower pressures of 20 to 80 GPa, a CSH system having the stoichiometry of $C_2S_2H_{10}$ [consisting of one $C_2H_6$ (ethane) molecule in hydrogen bonding with two $H_2S$ molecules], is most stable. It gives the major Raman shift peaks observed in experiment.

To determine the possible stoichiometry and stable crystal structures in a wide pressure range from 20 to 300 GPa, we used the evolutionary structure searching code *USPEX* [21] combined with the DFT code *VASP* [22] to perform systematic searches with various combinations of $C_iS_jH_k$ by varying the individual atom numbers *i, j, k* in a unit cell. We have mostly scanned the following range of combinations: $i$ = 1 to 5, $j$ = 1 to 5, and $k$ = 3 to 36 for discrete pressure points at 43, 100, 120, 150, and 250 GPa. Using the electronic formation enthalpy defined as $\Delta H = [H(C_iS_jH_k) - i \times H(C) - j \times H(S) - k \times H(H)]/(i + j + k)$ as the criterion, we have identified possible ternary compounds that are stable [$\Delta H$(eV/atom) < 0 similar to the so-called convex-hull analysis] in the searched pressure and stoichiometry range. Here the enthalpy [H(C), H(S), and H(H)] for each individual element of C, S, and H are calculated separately from their own stable phases at the corresponding pressure. In all of our DFT calculations, we have used the generalized gradient approximation (GGA) Perdew–Burke–Ernzerhof (PBE) exchange-correlation functional [23], although we also checked our calculations with the newly developed meta-GGA exchange-correlation (xc) functional SCANL [24] that is the orbital-free version of SCAN [25]. The latter xc functional SCANL better describes the $H_2$ dissociation under high pressure as demonstrated in a recent publication [26]. We find that SCANL calculations give qualitatively similar results to that of PBE calculations, although SCANL overall lowers the formation enthalpy of CSH compounds by ~10 to 20 meV/atom. For this reason, we only report and discuss PBE results in this letter.

As an example of our CSH-compound searches at 250 GPa (shown by Fig. S1 in Supplemental Materials [27]), the electronic formation enthalpy $\Delta H$ is plotted as a function of hydrogen atom number in a unit cell. These results show a general trend; that is, CSH compounds composed of one $CH_4$ molecule and integer numbers of $H_3S$ molecules tend to have lower formation enthalpy. Examples of such compounds are $C_1S_2H_{10}$ [$CH_4$ + $(H_3S)_2$], $C_1S_3H_{13}$ [$CH_4$ + $(H_3S)_3$], $C_1S_4H_{16}$ [$CH_4$ + $(H_3S)_4$], and $C_1S_5H_{19}$ [$CH_4$ + $(H_3S)_5$], in which the $CH_4$ molecule is a "guest" to the "host" $H_3S$ structure. This energetics observation of CSH compounds prompts us to look into their binary components of $C_iH_k$, $S_jH_k$, and $C_iS_j$ at the same pressure of 250 GPa. To that end, we have carried out similar evolutionary stoichiometry and structure searches for a variety of binary combinations. These binary searching results (shown by Figs. S2 and S3 in Supplemental Materials [27]) lead to the following conclusions: (1) the basic unit of $H_3S$ at its $Im\bar{3}m$ cubic structure gives the lowest formation enthalpy of $\Delta H_{H_3S} \approx -0.125$ eV/atom among all binary compounds we searched; (2) the basic unit of $C_2H_4$ has the lowest formation enthalpy [$\Delta H_{C_2H_4} \approx -0.062$ eV/atom] among $C_jH_k$ binary systems, even though it is twice as high as that of $H_3S$; and (3) the $C_iS_j$ binary compounds tend to be unstable ($\Delta H_{C_iS_j} > 0$) in this high pressure of 250 GPa. From the thermodynamic stability point of view, these results unambiguously suggest that a higher $H_3S$ concentration in a ternary CSH compound should drive the system to lower formation enthalpy (more stable). This leads us to explore the energetics and stability of carbon doping in $H_3S$ systems. It is noted that substitution of the sulfur atom in $H_3S$ by oxygen, phosphorus, and other elements had been studied previously [28], although it was not motivated from the energetics point of view.

To examine the formation enthalpy of carbon-doped $H_3S$ systems, we have varied the ratio of carbon to sulfur in a "super-cell" composed of integer numbers of $H_3S$ units in its *R3m* or *Im$\bar{3}$m* structure. Taking systems like $C_1S_9H_{30}$, $C_1S_{11}H_{36}$, $C_1S_{15}H_{48}$, $C_1S_{17}H_{54}$, $C_1S_{23}H_{72}$, $C_1S_{26}H_{81}$, and $C_1S_{35}H_{108}$ as examples, we replace one S atom by C in these $H_3S$ systems and relax them using VASP. We then calculated the formation enthalpy of these energetically optimized CSH compounds at the pressure range of 80 to 300 GPa. In order to add the zero-point energy (ZPE) difference ($\Delta E_{ZPE}$) to the electronic results of $\Delta H$, we performed the density-functional perturbation theory (DFPT) calculations for phonon spectra of these compounds as well as the corresponding individual elements. Namely, $\Delta E_{ZPE}(eV/atom) = [E_{ZPM}(C_iS_jH_k) - i \times E_{ZPM}(C) - j \times E_{ZPM}(S) - k \times E_{ZPM}(H)]/(i + j + k)$ is the averaged zero-point-motion (ZPM) contribution of ions to the total formation enthalpy of CSH systems. Finally, the total formation enthalpy ($\Delta H_{total} = \Delta H + \Delta E_{ZPE}$) versus pressure is plotted in Fig. 1 for the most-stable CSH compounds examined.

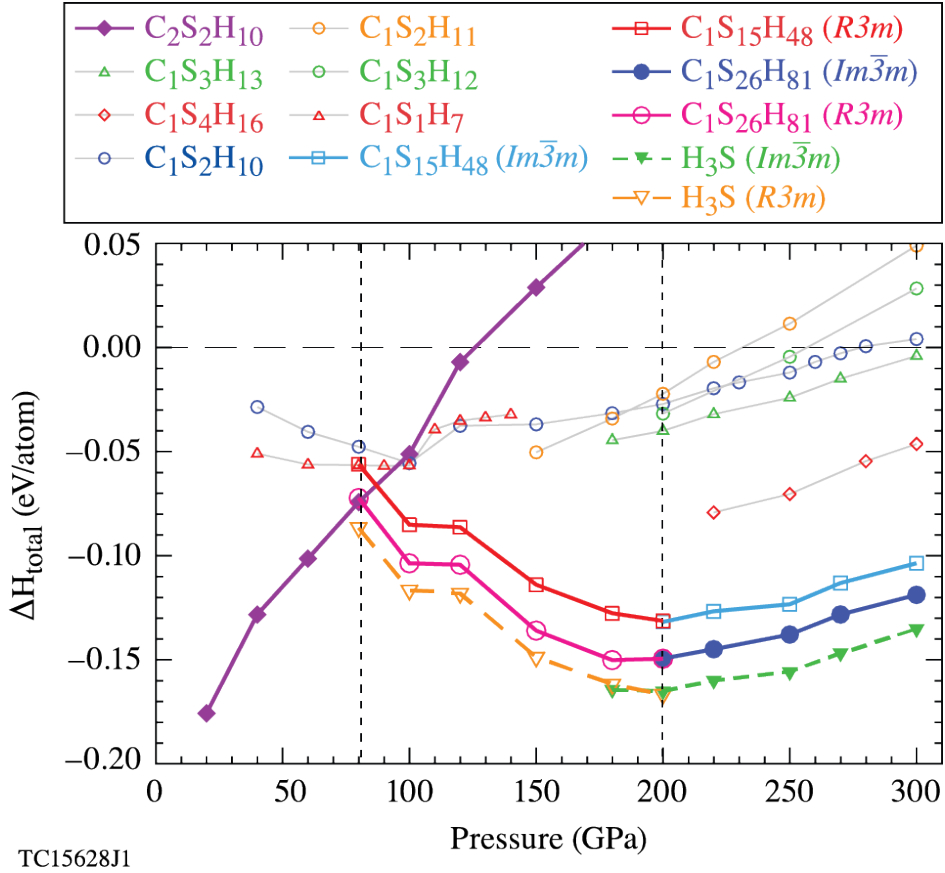

FIG. 1. The total formation enthalpy versus pressure for different CSH compounds in the pressure range of 20 to 300 GPa.

In Fig. 1 we have also included the formation enthalpy of $H_3S$ in its $R3m$ phase (80 to 180 GPa) and $Im\bar{3}m$ phase (200 to 300 GPa) for comparison. As one can see from Fig. 1 the $H_3S$ system has the lowest formation enthalpy for all the structures we examined so far in this pressure range (>80 GPa). Carbon-doped sulfur hydrides, varying from $C_1S_{35}H_{108}$ to $C_1S_9H_{30}$, are energetically second to the pure $H_3S$ system within an ~50-meV difference of $\Delta H_{total}$. Two such examples are plotted in Fig. 1: $C_1S_{15}H_{48}$ and $C_1S_{26}H_{81}$. Upon increasing the carbon doping level [defined as $d = N_C/(N_C + N_S)$ with $N_C$ and $N_S$ are the numbers of C and S atoms in the compound], the formation enthalpy of CSH compounds elevates from the pure $H_3S$. Nevertheless, these carbon-doped sulfur hydrides are both thermodynamically and dynamically stable since they all have $\Delta H_{total} < 0$ and no imaginary components in their phonon spectra. Once the carbon concentration increases in compounds like $C_1S_2H_{10}$, $C_1S_3H_{13}$, $C_1S_4H_{16}$, and other stoichiometry identified from structure searches, their total formation enthalpy approaches zero as shown in Fig. 1. The stoichiometry $C_1S_1H_7$ identified previously [19,20] is also stable in the pressure range of 40 to 150 GPa, but it has higher formation enthalpy than that of $H_3S$ in the *R3m* phase. At low pressures (<80 GPa), the CSH system takes the most-stable stoichiometry of $C_2S_2H_{10}$, composed of one $C_2H_6$ (ethane) molecule linked by hydrogen bonding with two $H_2S$ molecules (see Fig. S4 in Supplemental Materials [27]).

To examine if these identified carbon-doped sulfur hydrides exhibit any high-temperature superconductivity, we have employed the Allen–Dynes–McMillian equation [29] implemented in the DFT code *Quantum-Espresso* [30] to estimate the superconducting transition temperature $T_c$ for compounds varying from $C_1S_9H_{30}$ to $C_1S_{35}H_{108}$ in the pressure range of 150 to 300 GPa. Due to the difficulty of handling the large number of atoms of such CSH compounds (like a "supercell"), we have adopted the virtual crystal approximation (VCA) that has been used previously in literature to simulate doped systems [28]. Under VCA, the effective potential of the combined atom of "$C_dS_{1-d}$" takes a fraction superposition of carbon and sulfur pseudopotentials of the Hartwigsen–Goedecker–Hutter type [31], i.e., $V_{eff} = d \times V_C + (1-d) \times V_S$, with the doping level defined as $d = N_C/(N_C + N_S)$ for a given number of C and S atoms in the compound. The validity of the VCA method has been tested before [28]; we have also checked it with density-of-state calculations, which compare well with direct super-cell calculations. A dense $k$ mesh of 36 × 36 × 36 with a fine $q$ grid of 9 × 9 × 9 has been used in our $T_c$ calculations using VCA. The $T_c$ results from our VCA calculations for $C_1S_{15}H_{48}$ and $C_1S_{17}H_{54}$ are plotted in Fig. 2(a) as a function of pressure to compare with experimental measurements [18]. The $T_c$ range marked for the calculation data reflects the reasonable range of Coulomb repulsion constant $\mu^* = 0.10$ to $0.15$ used in the Allen–Dynes–McMillian equation. Overall, good agreement between calculation and experiment is reached with carbon-doped sulfur hydrides of $C_1S_{15}H_{48}$ and $C_1S_{17}H_{54}$ in the pressure range, which gives a room-temperature $T_c = 280\pm10$ K at ~270 GPa. Figure 2(b) plots the estimated $T_c$ variation as a function of carbon-doping level, corresponding to the stable CSH compounds ranging from $C_1S_9H_{30}$ to $C_1S_{35}H_{108}$ at 270 GPa. A peak of average $T_c \sim 290$ K appears at an *optimal* carbon-doping level of ~5.56% [corresponding to $C_1S_{17}H_{54}$]. In the CSH experiment [18] there was no intentional control of the carbon-doping level. We speculate that the experimentally synthesized CSH samples in the diamond-anvil cell (DAC) might contain a range of possible carbon-doping levels in these sulfur hydrides. However, any existing compounds around ~5%─6% carbon doping should dominantly contribute to the measured zero resistance. This is because the inverse of total resistance of the macroscopic sample is the "*algebraic sum*" of the inverse resistance of all components in the sample [according to Ohm's law for parallel resistors], i.e., $R_{\text{total}} = 1 \big/ \left( \frac{1}{R_1} + \frac{1}{R_2} + \frac{1}{R_3} + ... \right)$. Namely, if there is one compound superconducting ($R_i \approx 0$) in the sample, one measures zero resistance for the whole sample; the measured curve of resistance versus temperature will have only one "*jump*" that is determined by the highest $T_c$ of any superconducting compound in the sample. These features seemed to be in line with experimental measurements of the CSH system [18].

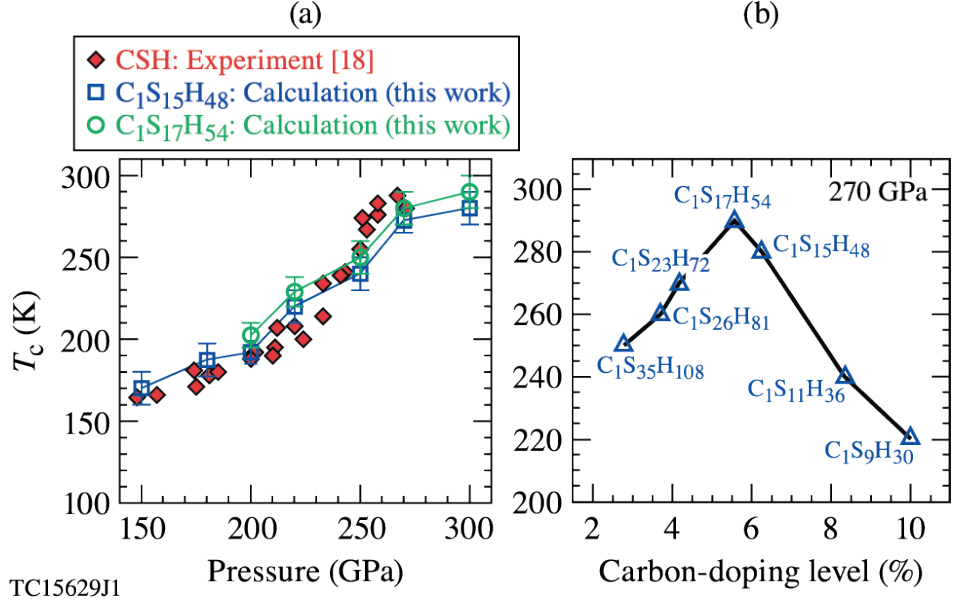

FIG. 2. (a) The calculated superconducting transition temperature $T_c$ versus pressure for compounds $C_1S_{15}H_{48}$ and $C_1S_{17}H_{54}$, compared to the CSH experiment [18]. (b) The average $T_c$ as a function of carbon doping level in different CSH compounds at the same pressure of 270 GPa.

To further analyze the physics mechanism behind the room-temperature superconductor of carbon-doped sulfur hydrides, we present the electronic and phonon density-of-state (DOS) calculation results in Fig. 3 for $C_1S_{17}H_{54}$ at pressures of 200 GPa and 270 GPa. It is noted that the carbon doping contributes more to the higher-frequency phonon mode due to its strong covalent bonding to hydrogen atoms. Figure 3(a) shows that the electronic DOS of $C_1S_{17}H_{54}$ is very similar for the two pressures, although high pressure makes the DOS peak at the Fermi surface slightly higher. The phonon DOS displayed by Fig. 3(b) indicates higher phonon frequencies appearing as pressure increases.

This trend leads to a significant enhancement of the average phonon frequency, changing from $\omega_{\ln} \sim 1010$ K at 200 GPa to $\omega_{\ln} \sim 1550$ K at 2700 GPa from shown by Fig. 3(c). With a strong electron–phonon coupling parameter of $\lambda \sim 2.5$ to 4.3 [also plotted in Fig. 3(c)], the carbon-doped sulfur hydrides like $C_1S_{17}H_{54}$ can have high-$T_c$ superconductivity in the room-temperature regime of ~270 GPa. The transition temperature $T_c$ is saturated at above ~270 GPa, which results from the drop in electron–phonon coupling despite the continuous increase of $\omega_{\ln}$.

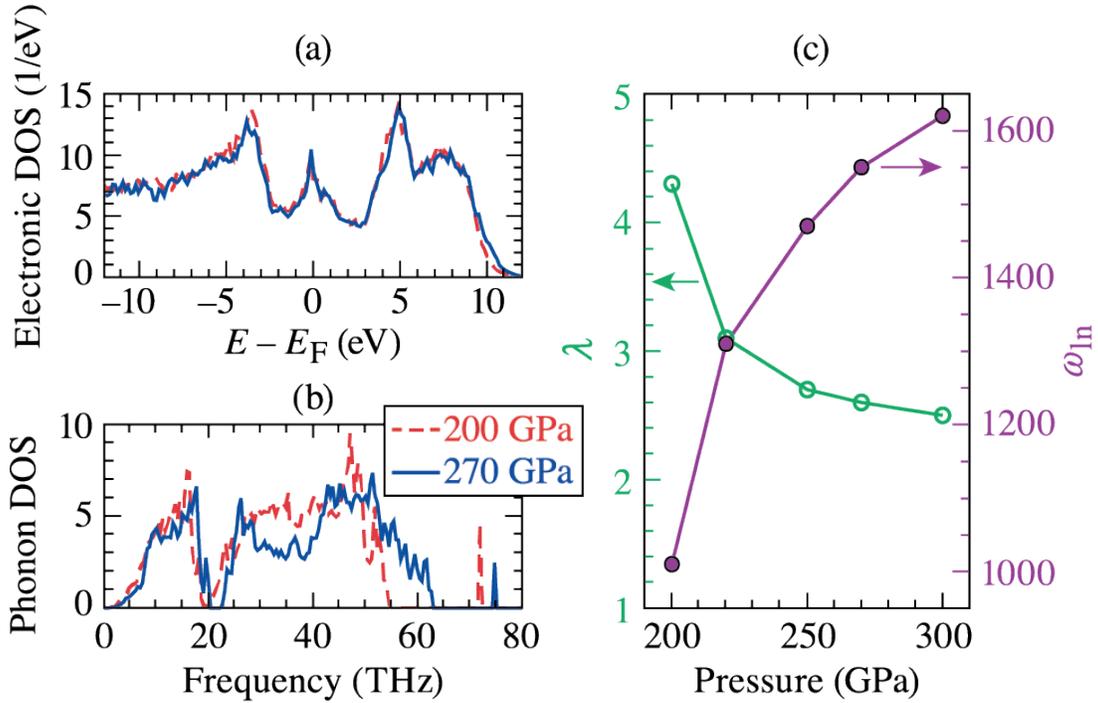

TC15631J1

FIG. 3. (a) The electronic density of state (DOS) for $C_1S_{17}H_{54}$ in the $Im\bar{3}m$ structure at pressures of 200 GPa (red dashed line) and 270 GPa (blue solid line). (b) Similar to (a) but for phonon DOS. (c) The calculated electron–phonon coupling parameter $\lambda$ and the average phonon frequency $\omega_{\ln}$ as a function of pressure for $C_1S_{17}H_{54}$.

Finally, we explore the possible composition of a CSH system experimentally synthesized at low pressures (<80 GPa), in which $C_2S_2H_{10}$ has been identified as the most-stable compound from our structure searches. To confirm if this stoichiometry is in line with experiment, we have performed Raman spectrum calculations of $C_2S_2H_{10}$ at ~37 GPa. The calculated Raman spectra [Fig. 4(b)] are compared with the CSH experiment measurement [18]. The $C_2S_2H_{10}$ structure (Fig. S4) consists of one ethane molecule ($C_2H_6$) surrounded by two $H_2S$ molecules in the unit cell. The two $H_2S$ molecules have different orientations, which results in distinct Raman shift signatures of a cluster of H-S-H modes at $v_3 = 100$ to 300 cm$^{-1}$, $v_2 = 600$ to 1200 cm$^{-1}$, and $v_1 = 2000$ to 2500 cm$^{-1}$. The overwhelming broad signal in the frequency range of $v = 1500$ to 2600 cm$^{-1}$ in the experiment [Fig. 4(a)] could be caused by the start of the $H_3S$ co-existing with the $C_2S_2H_{10}$ crystal in the DAC environment. This is supported by the Raman spectrum of $H_3S$ [black dashed line in Fig. 4(b)], in which a strong and broad peak shows up in this frequency range. The Raman peaks at $v = 3000$ to 3200 cm$^{-1}$ belong to the C-H modes of ethane molecule. We believe the peaks at $v_1 \approx 1500$ cm$^{-1}$ and $v_2 \approx 300$ to 600 cm$^{-1}$ are the fingerprints of the C-C vibrational and rotational modes of ethane in the synthesized $C_2S_2H_{10}$ crystal, even though the former is overwhelmed by the "*background*" signal of $H_3S$ in the experiment. Note that the calculated Raman spectra do not take into account any instrument resolution and broadening in experiment; therefore, the magnitude and width of the Raman peaks from calculations and measurements could vary. Nevertheless, the calculated locations and numbers of Raman peaks overall match the experimental measurements (although some small but noticeable frequency shifts are seen). Based on these comparisons, we conclude that $C_2S_2H_{10}$ should be the major material synthesized in experiment at low pressures, although $H_3S$ might start to coexist with $C_2S_2H_{10}$ at ~40 to 80 GPa. As the pressure increases, the major composition of CSH systems shifts to $H_3S$ and carbon-doped $H_3S$ compounds, while the latter gives rise to room-temperature superconductivity seen in both experiment and calculation. Additional peaks were observed in experiment beyond $\nu = 3500$ cm$^{-1}$. However, we speculated that these peaks should have resulted from the splitting of the hydrogen (H-H) peak upon interaction of a few hydrogen molecules with the CSH lattice. We inserted random hydrogen molecules in the periphery of a CSH unit cell and performed VASP QMD simulations at $T=300$K. Thereafter, we performed Raman calculations on this cell and verified this hypothesis.

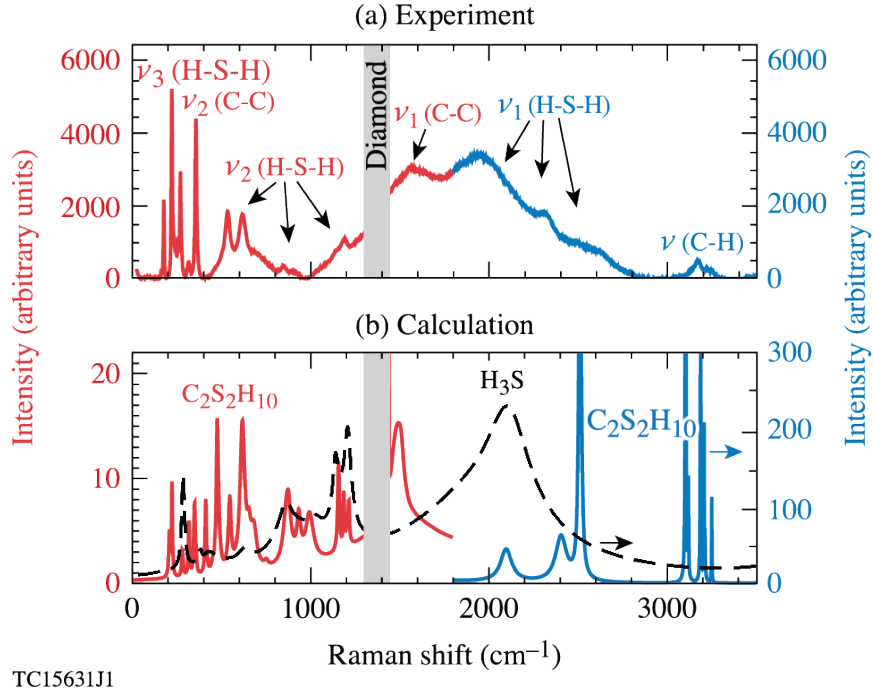

TC15631J1

FIG. 4. The Raman spectrum comparison between (a) the CSH experiment [18] and (b) our calculation of $C_2S_2H_{10}$ (solid red/blue lines) and $H_3S$ (dashed black line) at 37 GPa. Due to their relative magnitude difference, the calculated spectra of $C_2S_2H_{10}$ is split in two parts in (b) for clarity. Experimental instrumental resolution and broadening are not considered in calculations.

In summary, we have applied evolutionary structure and stoichiometry searches for the newly synthesized room-temperature superconducting phase of carbonaceous sulfur hydride. Using the total formation enthalpy ($\Delta H_{total}$) of CSH compounds, that possibly existed in the experimental sample, as the determining criterion we conclude that carbon-doped sulfur-hydride compounds like $C_1S_{15}H_{48}$ to $C_1S_{35}H_{108}$ in the cubic $Im\bar{3}m$ structure (>200 GPa) and *R3m* phase (80 to 200 GPa) are the most-stable entities, second to the pure $H_3S$ itself. The difference in $\Delta H_{total}$ between these CSH compounds and $H_3S$ is so small (<50 meV/atom) that they may co-exist with $H_3S$. Due to the replacement of small amount of S atoms by lighter carbon atoms, the logarithmic-averaged phonon frequency can be increased with pressure and modest electron–phonon coupling ($\lambda > 2.5$) results in these carbon-doped sulfur-hydride compounds. Our calculations show that an optimal carbon-doping level of $d \approx 5\%$ to 6% corresponds to compounds varying from $C_1S_{23}H_{72}$ to $C_1S_{15}H_{48}$ can lead to superconductivity with room-temperature $T_c$ at ~270 GPa. In contrast to the behavior of $H_3S$ [8,32], the slope of $dT_c/dP$ is positive for these identified CSH compounds. All of these findings seem to be in line with experimental observations. In addition, both the formation enthalpy and Raman spectrum calculations indicate that the CSH system is mainly composed of $C_2S_2H_{10}$ in a P1 structure (an ethane molecule in hydrogen bonding with two $H_2S$ molecules), which is again supported by the comparison of Raman spectrum with experiment. Given the demonstrated computational capabilities to handle ternary systems, we hope these results provide a stepping stone to guide future experiments toward the ultimate goal of making room-temperature superconductors at the most-accessible lower pressures ($P < 100$ GPa) or even at ambient conditions.

**Acknowledgment and Disclaimer**

This material is based upon work supported by the Department of Energy National Nuclear Security Administration under Award Number DE-NA0003856, the University of Rochester, and the New York State Energy Research and Development Authority. This work is partially supported by US National Science Foundation PHY Grant No. 1802964 for SXH and VVK.

This report was prepared as an account of work (for SXH, RP, and VVK) sponsored by an agency of the U.S. Government. Neither the U.S. Government nor any agency thereof, nor any of their employees, makes any warranty, express or implied, or assumes any legal liability or responsibility for the accuracy, completeness, or usefulness of any information, apparatus, product, or process disclosed, or represents that its use would not infringe privately owned rights. Reference herein to any specific commercial product, process, or service by trade name, trademark, manufacturer, or otherwise does not necessarily constitute or imply its endorsement, recommendation, or favoring by the U.S. Government or any agency thereof. The views and opinions of authors expressed herein do not necessarily state or reflect those of the U.S. Government or any agency thereof.

# Supplemental Material: Carbon-doped Sulfur Hydrides as *Room-Temperature* Superconductors at 270 GPa

S. X. Hu[1,2], R. Paul[1,2], V. V. Karasiev[1], R. P. Dias[2]

[1]Laboratory for Laser Energetics, University of Rochester,

250 East River Road, Rochester, NY 14623-1299, USA

[2] Department of Mechanical Engineering, University of Rochester,

250 East River Road, Rochester, NY 14623-1299, USA

To search for the stoichiometry and structure that are possibly stable in the CSH experiment under high pressures, we have applied the evolutionary structure searching code USPEX in combination with the density-functional theory (DFT) code VASP to perform systematic searches with various compounds of $C_iS_jH_k$, by varying their individual atom numbers $i, j, k$ in a unit cell. We have mostly scanned the following range of combinations: $i = 1 - 4$, $j = 1 - 4$, and $k = 3 - 36$ for discrete pressure points at 43, 100, 120, 150, and 250 GPa. Using the electronic formation enthalpy defined as $\Delta H = [H(C_iS_jH_k) - i \times H(C) - j \times H(S) - k \times H(H)]/(i + j + k)$ as the criterion, we have identified possible ternary compounds to be thermodynamically stable if they have a negative formation enthalpy [ $\Delta H(eV/atom) < 0$]. Here the enthalpy $[H(C), H(S), and\ H(H)]$ for each individual element of C, S, and H are calculated separately from their own stable phases at the corresponding pressure. As an example, Fig. S1 shows $\Delta H$ as a function of hydrogen atom number in a compound of $C_iS_jH_k$ at 250 GPa:

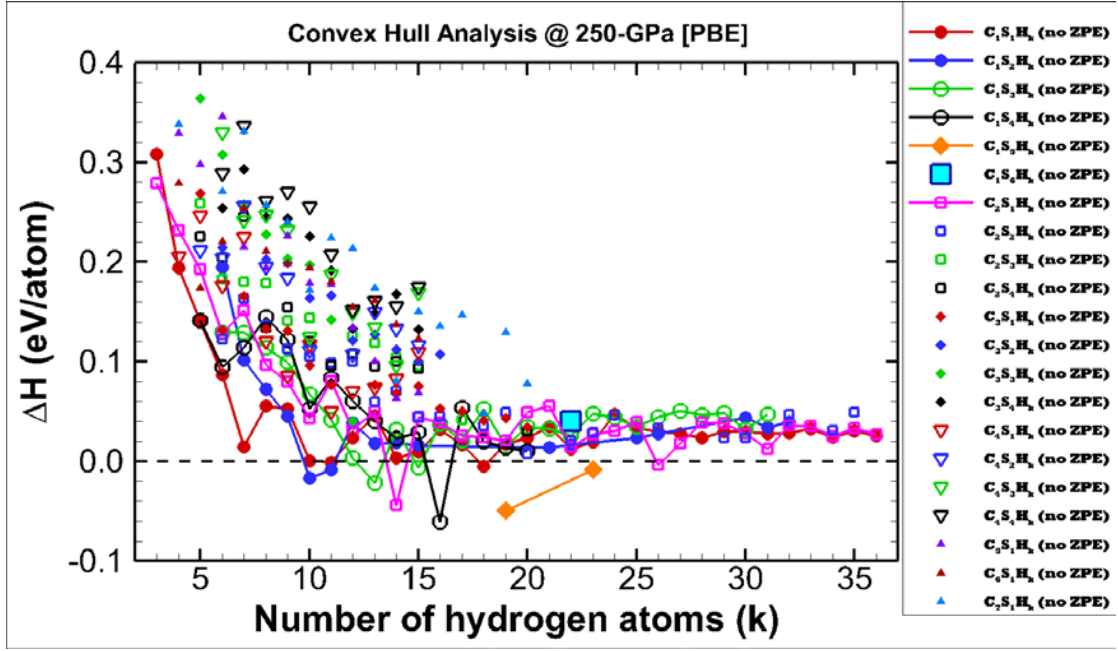

Figure S1. (Color online) The electronic formation enthalpy $\Delta H$ is plotted as a function of hydrogen atom number $k$ in different compounds of $C_iS_jH_k$ at 250 GPa, for most combinations of $i = 1-4$ and $j = 1-4$ for C and S atoms, respectively. A few cases of $j = 5, 6$ are also shown.

The convex hull analyses for the binary sub-systems of $C_iH_k$, $S_jH_k$, and $C_iS_j$ at the same pressure of 250 GPa are shown by Figs. S2 and S3. It is shown by Fig. S2 that the two basic binary units of $C_2H_4$ and $S_1H_3$ are the most stable ones at this pressure. Figure S3 indicates the binaries composed by C and S atoms only are not stable for the combinations explored. In general, they have much large positive formation enthalpy ($\Delta H > +0.20$ eV/atom). Namely, without any hydrogen atom involved in the compounds, carbon and sulfur atoms are unlikely forming any stable compounds at this high pressure.

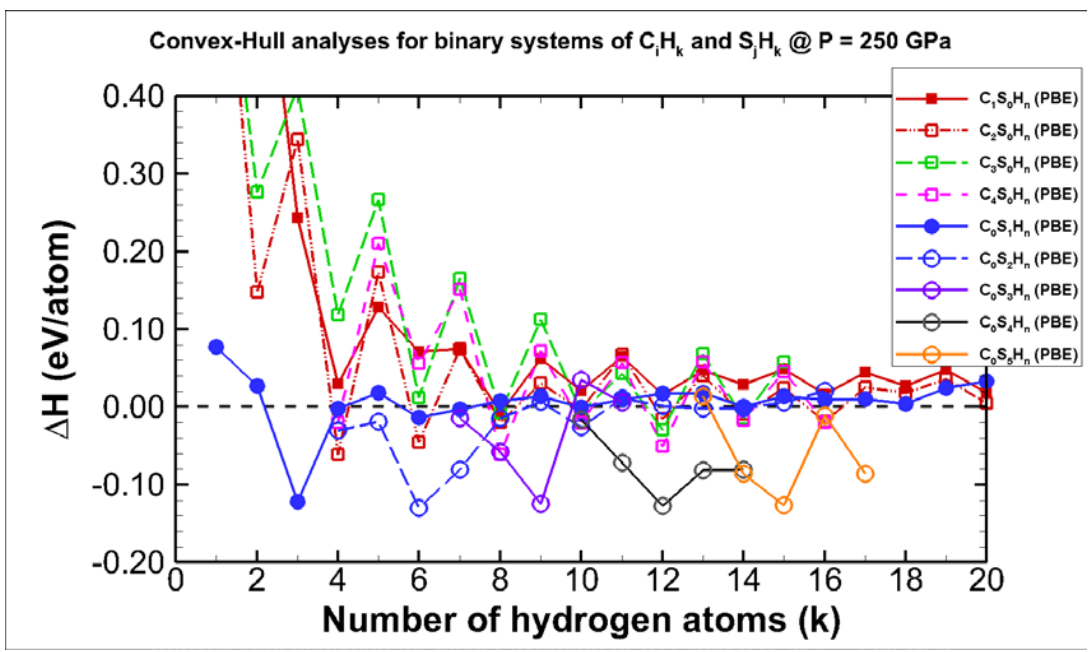

Figure S2. (Color online) The electronic formation enthalpy $\Delta H$ is plotted as a function of hydrogen atom number $k$ for binary sub-systems of $C_iH_k$ and $S_jH_k$ at 250 GPa.

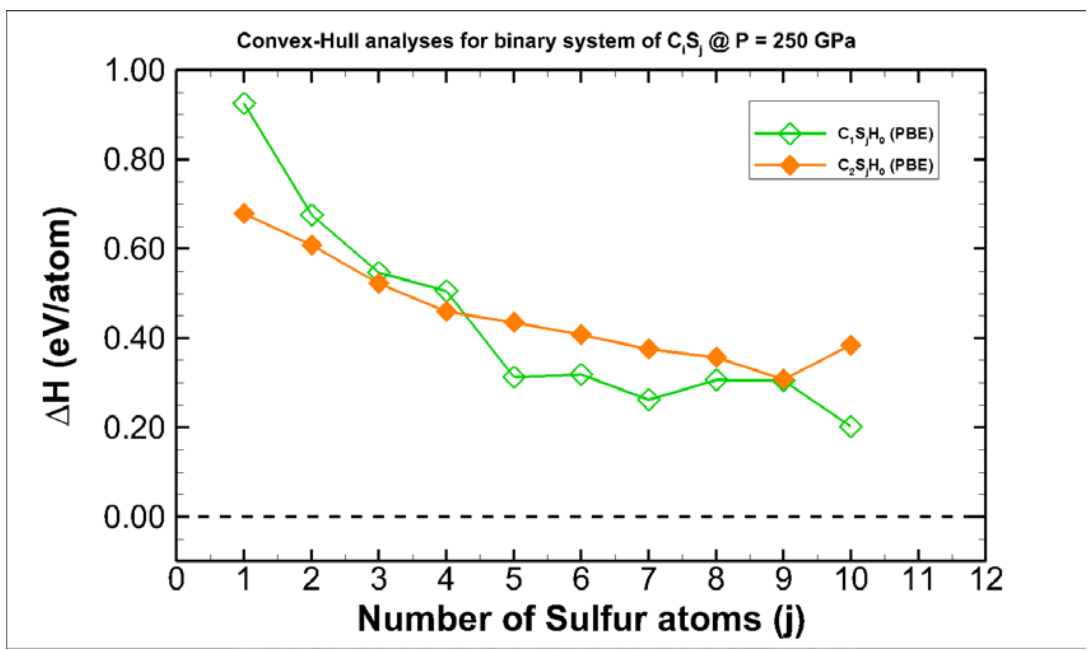

Figure S3. (Color online) The electronic formation enthalpy $\Delta H$ is plotted as a function of sulfur atom number $j$ for binary sub-systems of $C_i S_j$ at 250 GPa.

Figure S4 shows the structure of $C_2S_2H_{10}$ that is the most stable stoichiometry of CSH system at relatively low pressures of $20-80$ GPa. It consists of one ethane ($C_2H_6$) molecule covalently bonding with two $H_2S$ molecules in a unit cell, having the *P1* symmetry. The two $H_2S$ molecules have different orientations. At 40 GPa the unit cell for $C_2S_2H_{10}$ has the following lattice parameters: $a = 3.126$ Å, $b = 5.150$ Å, $c = 5.007$ Å and $\alpha = 62.82°, \beta = 89.36°, \gamma = 95.97°$, from the USPEX search and VASP optimization.

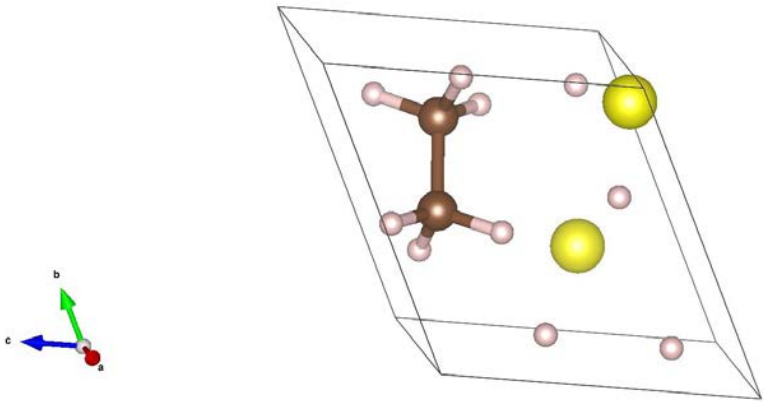

Figure S4. (Color online) The structure of $C_2S_2H_{10}$ that is the most stable stoichiometry of CSH system at relatively low pressures of $20-80$ GPa.